\documentclass[aps,prl,twocolumn,superscriptaddress,showpacs]{revtex4-1}

\usepackage{graphicx}
\usepackage{amsmath}
\usepackage{float}

\begin{document}
\title{Optical Polarization M\"obius Strips and Points of Purely Transverse Spin Density}

\author{Thomas Bauer}
\affiliation{Max Planck Institute for the Science of Light, Guenther-Scharowsky-Str. 1, D-91058 Erlangen, Germany}
\affiliation{Institute of Optics, Information and Photonics, University Erlangen-Nuremberg, Staudtstr. 7/B2, D-91058 Erlangen, Germany}
\author{Martin Neugebauer}
\affiliation{Max Planck Institute for the Science of Light, Guenther-Scharowsky-Str. 1, D-91058 Erlangen, Germany}
\affiliation{Institute of Optics, Information and Photonics, University Erlangen-Nuremberg, Staudtstr. 7/B2, D-91058 Erlangen, Germany}
\author{Gerd Leuchs}
\affiliation{Max Planck Institute for the Science of Light, Guenther-Scharowsky-Str. 1, D-91058 Erlangen, Germany}
\affiliation{Institute of Optics, Information and Photonics, University Erlangen-Nuremberg, Staudtstr. 7/B2, D-91058 Erlangen, Germany}
\affiliation{Department of Physics, University of Ottawa, 25 Templeton, Ottawa, Ontario K1N 6N5 Canada}
\author{Peter Banzer}
\email[]{peter.banzer@mpl.mpg.de}
\homepage[]{http://www.mpl.mpg.de/}
\affiliation{Max Planck Institute for the Science of Light, Guenther-Scharowsky-Str. 1, D-91058 Erlangen, Germany}
\affiliation{Institute of Optics, Information and Photonics, University Erlangen-Nuremberg, Staudtstr. 7/B2, D-91058 Erlangen, Germany}
\affiliation{Department of Physics, University of Ottawa, 25 Templeton, Ottawa, Ontario K1N 6N5 Canada}

\date{\today}

\begin{abstract}
Tightly focused light beams can exhibit electric fields spinning around any axis including the one transverse to the beams' propagation direction. At certain focal positions, the corresponding local polarization ellipse can degenerate into a perfect circle, representing a point of circular polarization, or C-point. We consider the most fundamental case of a linearly polarized Gaussian beam, where -- upon tight focusing -- those C-points created by transversely spinning fields can form the center of 3D optical polarization topologies when choosing the plane of observation appropriately. Due to the high symmetry of the focal field, these polarization topologies exhibit non trivial structures similar to M\"obius strips. We use a direct physical measure to find C-points with an arbitrarily oriented spinning axis of the electric field and experimentally investigate the fully three-dimensional polarization topologies surrounding these C-points by exploiting an amplitude and phase reconstruction technique. 
\end{abstract}


\maketitle

\begin{figure*}[htbp]
\centerline{\includegraphics[width=2\columnwidth]{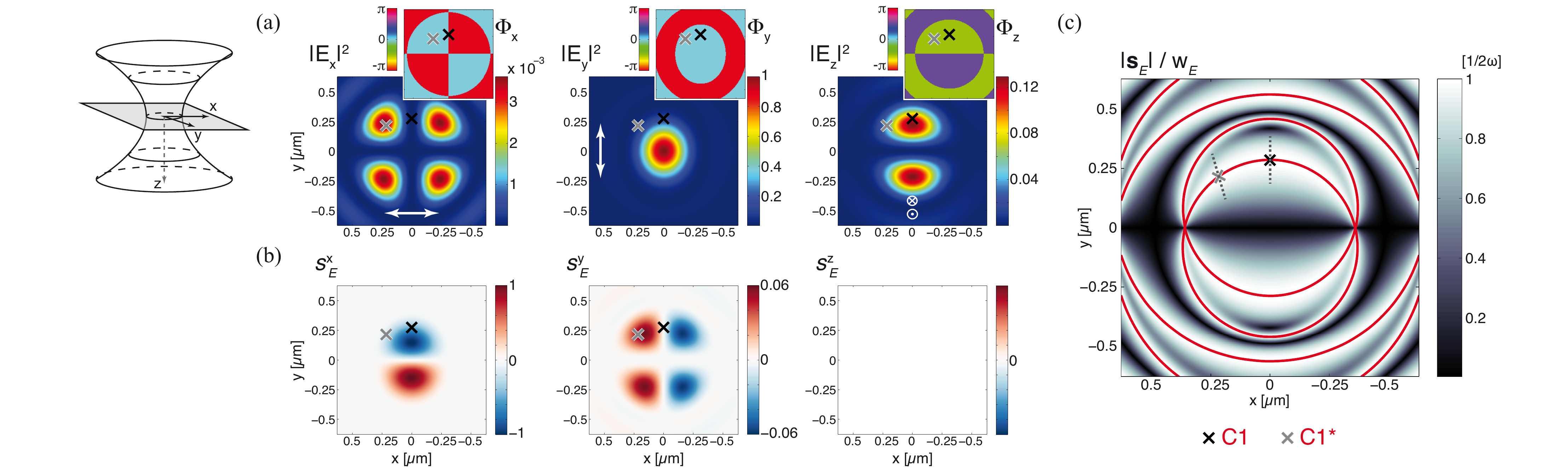}}
\caption{\label{fig1}(color). (a) The theoretically calculated components of the focal electric energy density distribution $\left|E_{\text{x}}\right|^{2}$, $\left|E_{\text{y}}\right|^{2}$ and $\left|E_{\text{z}}\right|^{2}$ of a tightly focused linearly $y$-polarized Gaussian beam, normalized to the maximum electric energy density. Corresponding phase distributions are plotted as insets. (b) The distributions of the two non-zero components of the transverse spin density $s_{E}^{x}$ and $s_{E}^{y}$ in the focal plane (normalized to the maximum value of $s_{E}^{x}$). Due to the symmetry of the light field, $s_{E}^{z}$ is identical zero across the whole focal plane. (c) Focal distribution of $\left|\textbf{s}_{E}\right|$ normalized to the local electric energy density $w_E$. The red solid lines correspond to the maximum value $\frac{1}{2 \omega}$. The black and gray markers in all distributions correspond to the considered C-points C1 on the $y$-axis and C1$^{*}$ on the bisector of the positive $x$- and $y$-axis, while the gray dashed lines show the principal plane of their polarization circle.}
\end{figure*}

\begin{figure*}[htbp]
\centerline{\includegraphics[width=2\columnwidth]{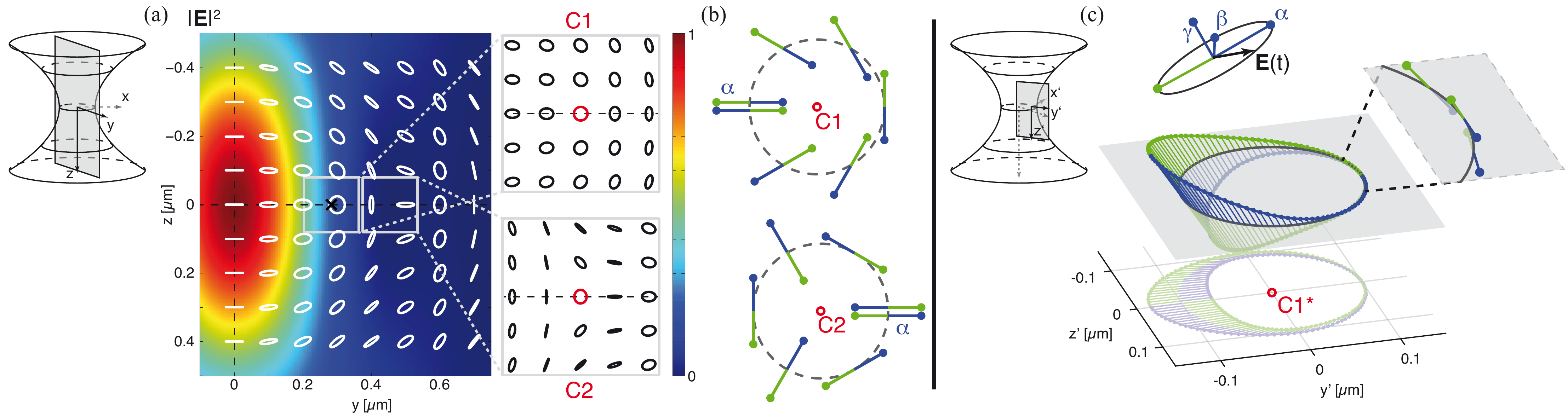}}
\caption{\label{fig2}(color). (a) The electric energy density distribution in the $yz$-plane, superimposed by the local polarization ellipses in white. Details in the vicinity of the first two C-points C1 and C2 along the $y$-axis are shown as insets. (b) Trace of the major axis of the polarization ellipse around the C-points marked in red in the insets in (a). (c) Arising optical polarization topology when tracing around C1$^{*}$ (in the principal plane of its polarization ellipse with local coordinates $x', y', z'=z$) with a trace radius of $100 \text{ nm}$. The occurring weak $x'$-component of the electric field is magnified 4 times to show the orientation of the major axis of the polarization ellipse more distinctly. The definitions of the axes in an arbitrary polarization ellipse are shown as inset. The magnified part of the trace shows the major axis rotating into the principal plane and pointing along the trace direction. }
\end{figure*}

\textit{Introduction.}\textemdash 
Structured light fields represent important tools in modern optics due to their multitude of different applications, for example in microscopy \cite{Quabis2000,Youngworth2000,Rittweger2009}, nano-optics \cite{Zuchner2008,Sancho-Parramon2012,Neugebauer2014,Wozniak2015} and optical trapping \cite{Dholakia2011,Taylor2015}. In general, the spatial tailoring of amplitude, phase and polarization of paraxial or non-paraxial light beams can lead to interesting and sometimes complex topological structures of the electric field such as phase vortices \cite{Nye1974,Allen1992}, optical knots \cite{Leach2004,Dennis2010} and optical polarization M\"obius strips \cite{Freund2005,Dennis2011,Bauer2015}. The latter are linked to the presence of so-called C-points or C-lines in the field distribution, i.e.\ positions, where the local polarization ellipse degenerates to a circle and the field is circularly polarized \cite{Nye1987}. While polarization M\"obius strips have been predicted a decade ago and investigated ever since \cite{Freund2010}, only recent advances in nano-optics, in particular 3D amplitude and phase reconstruction techniques at the nanoscale \cite{Bauer2013}, enabled their experimental verification \cite{Bauer2015}. This was achieved by tightly focusing a light beam with spatially varying polarization spanning the full Poincar\'{e} sphere \cite{Beckley2010,Galvez2012,Cardano2013}. In the focal plane of such a full Poincar\'{e} beam, this yields a complex fully vectorial field structure, including a C-point on the optical axis with the electric field spinning around said (longitudinal) axis. Tracing the major axis of the polarization ellipse around this C-point allowed for revealing optical polarization M\"obius strips hidden in the focal polarization distribution \cite{Bauer2015}.\\
In more general field distributions, C-points with an arbitrarily oriented spinning axis of the electric field might be observed. This includes the special case of a spinning axis perpendicular to the optical axis \cite{Nye1987}, which indicates the presence of purely transverse spin angular momentum (tSAM) \cite{Aiello2015-2}. While the rising interest in this intriguing polarization component is strongly linked to its occurrence in highly confined fields within guided modes and plasmons \cite{Bliokh2012,Rodriguez-Fortuno2013,Petersen2014}, it was shown that tSAM is also naturally present in many focusing scenarios \cite{Yang2011,Banzer2013,Neugebauer2014,Neugebauer2015,Aiello2015-2}. As an example, even an ordinary beam such as a linearly polarized fundamental Gaussian beam exhibits transversely spinning fields when tightly focused \cite{Richards1959,Neugebauer2014,Neugebauer2015}. In this case, the transverse ($\textbf{E}_{\bot}$) and longitudinal ($E_{z}$) electric field components exhibit in the focal plane a $\pi/2$ phase difference with respect to each other, and positions where both field components have the same amplitude represent C-points that are linked to purely tSAM. This raises the question, whether those focal C-points of a Gaussian beam also exhibit complex polarization topologies. \\
In this paper, we investigate the occurrence of M\"obius-like optical polarization topologies in the focal field distribution of a tightly focused linearly polarized Gaussian beam. Furthermore, we show the generation of optical polarization M\"obius strips around C-points with non-zero transverse spin in the experimental realization of the mentioned field configuration using a nanointerferometric amplitude and phase reconstruction technique \cite{Bauer2013}.
\\
\textit{C-points, transverse spin and optical polarization M\"obius strips.}\textemdash
When examining polarization topologies in fully vectorial and highly confined structured light fields, the electric field $\mathbf{E}$ at each point $\mathbf{r}$ is in general oscillating in a plane oriented arbitrarily in space. This means, that the local field can be described by a polarization ellipse with its major axis $\boldsymbol{\alpha} (\mathbf{r})$, minor axis $\boldsymbol{\beta} (\mathbf{r})$ and normal vector $\boldsymbol{\gamma} (\mathbf{r})$ defined in 3D-space by \cite{Berry2004}
\begin{align}\label{eqn_ellipse}
\boldsymbol{\alpha} &= \frac{1}{| \sqrt{\mathbf{E \cdot E}}|} \operatorname{Re} \left( \mathbf{E}^\ast \sqrt{\mathbf{E \cdot E}} \right) , \nonumber\\
\boldsymbol{\beta} &= \frac{1}{| \sqrt{\mathbf{E \cdot E}}|} \operatorname{Im} \left( \mathbf{E}^\ast \sqrt{\mathbf{E \cdot E}} \right) ,\\
\boldsymbol{\gamma} &= \operatorname{Im} \left( \mathbf{E}^\ast \times \mathbf{E} \right) , \nonumber
\end{align}
with $\mathbf{E}^\ast$ denoting the complex conjugate of the field. Thus, the field is oscillating in the plane spanned by $\boldsymbol{\alpha}$ and $\boldsymbol{\beta}$. The normal vector $\boldsymbol{\gamma}$ is, in this definition, directly proportional to the electric part of the spin density \cite{Berry2001,Bliokh2014,Aiello2015}
\begin{align}\label{eqn_S1}
\textbf{s}_{E}\left(x,y,z\right)= \frac{\epsilon_{0}}{4 \omega}\operatorname{Im} \left(\textbf{E}^{\ast}\times\textbf{E}\right) \text{,}
\end{align} 
with $\epsilon_0$ the permittivity of free space and $\omega$ the angular frequency of the monochromatic light field. Since the electric spin density specifies the orientation and sense of the spinning axis of the local electric field and represents a physically measurable quantity \cite{Bliokh2014,Neugebauer2015}, an elegant and straight forward method to determine points of arbitrarily oriented circular polarization (C-points) in a vectorial light field is to normalize $\textbf{s}_{E}$ with the energy density of the electric field, $w_{E}=\frac{\epsilon_{0}}{2}\left(\left|E_{x}\right|^2+\left|E_{y}\right|^2+\left|E_{z}\right|^2\right)$ \cite{Berry2001}. This amplitude-independent measure is maximized when the polarization ellipse degenerates to a circle, and we thus can define a simple requirement for C-points:
\begin{align}\label{eqn_S2}
\frac{\left| \textbf{s}_{E} \right|}{w_{E}}=
\frac{1}{2 \omega} \text{.}
\end{align}
As discussed before, the polarization ellipse can show intriguing topological patterns around such points in space. By choosing a plane of observation containing the C-point and with the normal vector of that plane parallel to the spinning axis $\boldsymbol{\gamma}$ of the electric field at this C-point, fundamental polarization topologies with a topological index of $\pm 1/2$ (in contrast to the integer numbered topological index of phase vortices) can be revealed \cite{Nye1987,Berry1977}. This half integer index is allowed, since the orientation of a (polarization) ellipse is indistinguishable under a rotation by $\pi$ \cite{Dennis2011}.\\
Within the aforementioned formalism, we investigate the field distribution of a tightly focused $y$-polarized Gaussian beam propagating along $z$. This scenario corresponds to one of the most fundamental beam configurations in optics labs and is used in many experimental studies. Figure \ref{fig1}(a) depicts the electric energy density and phase (insets) distributions of all three Cartesian components in the focal plane \cite{Richards1959,L.NovotnyandB.Hecht2006} for a numerical aperture of $0.9$, a wavelength of $\lambda=530 \text{ nm}$ and a fill-factor of the focusing aperture of $w_0/f=1.21$. 
The dominant field component $\left|E_{y}\right|^{2}$ has its maximum on the optical axis, while the longitudinal field $\left|E_{z}\right|^{2}$ shows a two-lobe structure, elongating the focal spot along the $y$-axis \cite{Richards1959,Dorn2003}. The crossed polarization component $\left|E_{x}\right|^{2}$, occurring analog to the longitudinal field component due to a rotation of the polarization components upon focusing, exhibits the characteristic four-lobe structure observed in cross-polarization measurements in many microscopy studies \cite{Novotny2001} and referred to as depolarization \cite{Bahlmann2000}. Considering the relative phases between the field components, we see that the longitudinal component is $\pm\pi/2$ out of phase with respect to the in-phase transverse components, indicating purely transversely spinning fields throughout the focal plane. This can also be seen in the components of the electric spin density (see Fig.~\ref{fig1}(b) and Ref.~\cite{Neugebauer2015}) and is a feature exhibited by many fundamental beams after tight focusing. The strongest and, therefore, dominant component of the electric spin density is $s_{E}^{x}$, with $s_{E}^{y}$ being weaker by one order of magnitude. As expected, the longitudinal component $s_{E}^{z}$ is identical zero, which implies that points of circular polarization in this case can only be linked to tSAM. In order to determine those C-points of purely transverse spin, the ratio $\left|\textbf{s}_{E}\right|/w_{E}$ is plotted in Fig.~\ref{fig1}(c). The red solid lines mark the C-lines, or equivalently, lines along which  Eq.~(\ref{eqn_S2}) is fulfilled. It is worth mentioning, that in this theoretically calculated field distribution, C-points can only be found in the actual focal plane of the linearly polarized tightly focused Gaussian beam. This is also shown in Fig.~\ref{fig2}(a), where the polarization ellipses are plotted as solid white lines on a cross-section of the electric energy density in the $yz$-plane for $x=0$. Due to the different phase velocities for the transverse and longitudinal electric field components of the tightly focused beam, all polarization ellipses outside the focal plane form ellipses, which in the far-field ($z \gg \lambda$) have to transform to the initially linear polarization. It can be seen in Figs.~\ref{fig1}(c) and \ref{fig2}(a) that along the $y$-axis in the focal plane additional C-points can be found further away from the optical axis.\\
In the following, we concentrate on three specific C-points and investigate their polarization topologies in appropriately selected planes of observation. First, we examine a C-point on the $y$-axis (C1), approximately $275 \text{ nm}$ away from the optical axis (see black crosses in Figs.~\ref{fig1}(a-c)), and look at the local polarization ellipses in its principal plane (here, the $yz$-plane).
All electric field vectors and, therefore, also the local polarization ellipses in this plane are in-plane only, since $E_{x}(0,y,z)=0$ (see also Fig.~\ref{fig1}(a)). Thus, we can use 2D polarization topologies to characterize the points of circular polarization in the chosen meridional plane of observation. With the terminology developed in Refs.~\cite{Berry1977,Nye1987}, C1 represents a lemon-type polarization topology. The major axis of the polarization ellipse rotates clockwise when traced clockwise around the central C-point and it performs a rotation of $\pi$ (see the upper part of Fig.~\ref{fig2}(b)). Thus, the topological index of C1 is $+1/2$. In contrast, the next C-point (C2) on the $y$-axis at a distance of approximately $455 \text{ nm}$ shows a star-type polarization topology (see the lower part of Fig.~\ref{fig2}(b)), associated with an index of $-1/2$. This alternation of the two different planar topologies continues when moving further away from the optical axis.\\
\begin{figure*}[htbp]
	\centerline{\includegraphics[width=2\columnwidth]{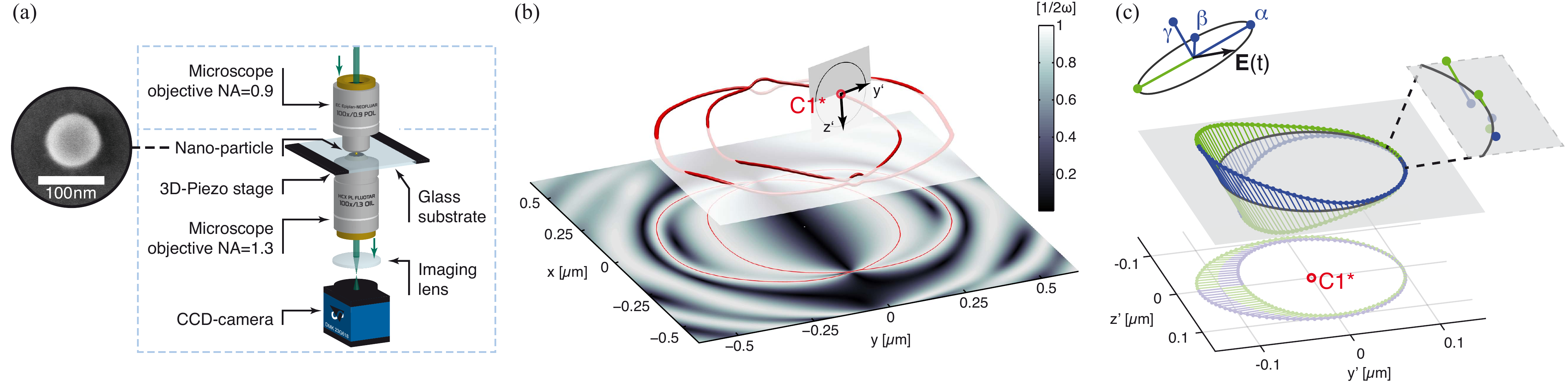}}
	\caption{\label{fig3}(color). (a) Sketch of the experimental setup for the reconstruction of the generated highly confined field distribution. An SEM image of the employed gold nanoprobe is shown as inset. (b) The experimentally reconstructed focal distribution of $\left|\textbf{s}_{E}\right| / w_E $ and the projections of the two innermost C-lines. The lower part depicts the projection of these C-lines onto the focal plane and the corresponding distribution of $\left|\textbf{s}_{E}\right| / w_E $.  The red lines in the upper part of (b) depict the 3D trajectories of both C-lines, which are crossing the focal plane (transparent white plane) repeatedly. The marked C-point C1$^{*}$ corresponds to the one considered in (c). (c) Optical polarization M\"obius strip with one half-twist, generated by tracing the major axis of the polarization ellipse around the C-point on the bisector of the positive $x$- and $y$-axis in the plane normal to its local spin vector, marked in gray. The trace radius was chosen to be $100 \text{ nm}$. The magnified part of the trace shows that the major axis is, due to slight phase aberrations, not at the same time parallel to the principal plane and pointing along the trace direction. The occurring weak $x'$-component of the electric field is magnified 4 times to show the half-twist of the major axis of the polarization ellipse more distinctly.}
\end{figure*}
Despite these sign changes of the topological index, the general structure of the occurring polarization topology and its local electric fields is limited to the $yz$-plane due to the missing out-of-plane electric field component. This means, that no 3D topologies can be observed in the $yz$-plane, which is the plane of symmetry of the overall focusing geometry. However, the symmetry can be broken by investigating a C-point off the $y$-axis and choosing the principal plane of its polarization ellipse as plane of observation. Since along a C-line, the topological index of the polarization singularity is conserved \cite{Nye1987}, we examine exemplarily the C-point on the bisector of the positive $x$- and $y$-axis closest to the optical axis (C1$^{*}$). Observing the field structure around this point in the coordinates $x', y', z'$ (with $z'=z$) of its principal plane marked with a gray dashed line in Fig.~\ref{fig1}(c), the electric field in the surrounding of the C-point exhibits a 3D field configuration due to a non-zero $E_{x'}$-component, see also Fig.~\ref{fig1}(a). This 3D field causes the major axis of the polarization ellipses around the C-point to tilt out of plane. As a result, the previously two-dimensional ellipse field around C1 is transformed into a 3D structure around C1$^{*}$ with a topology similar to an optical polarization M\"obius strip (see Fig.~\ref{fig2}(c)). To highlight the out-of-plane orientation of the polarization ellipses, the $x'$-components of the plotted major axes are magnified by a factor of 4. The position of the discontinuity present in the trace of the major polarization axis is given by an arbitrary choice of the offset phase of the field (as in the planar case, also seen in the projection onto the $y'z'$-plane in Fig.~\ref{fig2}(c)). It is important to mention again, that all components of the electromagnetic field are fully continuous at this point, only the major axis of the polarization ellipse exhibits a discontinuity. In contrast to the optical polarization M\"obius strips investigated in Refs. \cite{Freund2005,Dennis2011,Bauer2015}, the major axis of the polarization ellipse coincides at one position along the chosen trace with the direction of the trace around the C-point (for C1$^{*}$: $z'=0$, $y'=0.1\text{ $\mu$m}$, see magnified part of Fig.~\ref{fig2}(c)) due to the purely tSAM in the whole focal plane. In other words, the major axis of the polarization ellipse is tangential to the trace at this point. This special orientation prevents the determination of the actual twist-number of the major polarization axis when traced around the C-point, rendering this case different from a generic M\"obius strip. However, as will be shown later on, only slight aberrations in the focal field distribution, for example due to experimentally unavoidable phase aberrations, might lead to a distribution exhibiting not only purely tSAM. This implies, that the major axis of the polarization ellipse is not tangential to the trace anymore, and even at points with locally purely tSAM, the slight aberrations of the focal field will lead to a M\"obius topology.
\\
\textit{Experimental approach and results.}\textemdash 
To verify this theoretical prediction, the experimental setup sketched in Fig.~\ref{fig3}(a) \cite{Banzer2010a} was used. 
The incoming linearly $y$-polarized Gaussian beam is tightly focused by a first microscope objective (MO) with a numerical aperture (NA) of 0.9. A single spherical gold nanoparticle (radius $r=42\text{ nm}$), acting as a local nanoprobe, is immobilized on a glass substrate and scanned through the focal plane. A second MO (immersion type, $\text{NA}=1.3$) collects the transmitted field, including the light scattered off the particle in the forward direction. For each position of the nanoprobe relative to the optical axis, back-focal-plane images of the second MO  are acquired, corresponding to the angular spectrum transmitted into the forward direction. This experimental data can be used to reconstruct the full vectorial focal field distribution following the technique introduced in \cite{Bauer2013}, since amplitude and phase information are encoded in the angular interference between the transmitted beam and the scattered light. More details about the experimentally measured scan data and the resulting focal field distribution can be found in the Supplementary Material \footnote{See Supplemental Material at [URL will be inserted by publisher] for details about the experimentally measured interferometric data and reconstructed focal field distribution.}.\\
Figure \ref{fig3}(b) illustrates the experimentally achieved distribution of $\left|\textbf{s}_{E}\right|/w_{E}$, calculated from the reconstructed fully vectorial electric field distribution. Due to small experimental deviations from the theoretically expected field distribution, the experimentally reconstructed C-lines are located not only in the focal plane, but cross it repeatedly. These deviations from the theoretically expected field pattern can be explained by small aberrations of the incoming beam and the microscope objective. Tracing the major axis of the polarization ellipse around the C-point at the same bisector used in Fig.~\ref{fig2}(c) (gray crosses in Fig.~\ref{fig1}) in the principal plane of its polarization ellipse results in the optical polarization M\"obius strip with one half-twist as depicted in Fig.~\ref{fig3}(c). This C-point is not a point of purely tSAM in the experimental case due to the experimentally present phase aberrations, and thus leads to the shown M\"obius strip in agreement with the theoretical predictions. Also here, the out-of-plane component of the field was magnified by a factor of 4 to highlight the twist that can be seen in the magnified inset in Fig.~\ref{fig3}(c). To confirm that also C-points with purely tSAM exhibit these polarization topologies, the major axis of the polarization ellipse was additionally traced around such a C-point, and the obtained polarization M\"obius strip can be seen in the Supplementary Material. This demonstration of an optical polarization M\"obius strip confirms the occurrence of this topological structure even in the basic scenario of a tightly focused linearly polarized beam.
\\
\textit{Conclusion.}\textemdash 
We investigated optical polarization topologies in a tightly focused linearly polarized Gaussian beam and showed that a special case of a M\"obius-like topology, directly linked to the purely transverse spin density of the light field and the associated C-lines in the focal plane, can be observed. By defining an appropriate plane of observation, we were able to trace the existing polarization topology around a C-point with a purely transverse spinning axis of the electric field. This M\"obius-like topology is present in an ideal beam at all C-points located off the beam's symmetry axes. As experimental verification of the realized polarization topologies, we applied an interferometric nanoprobing technique, where slight aberrations of the focal field lead to the observation of generic optical polarization M\"obius strips.\\
Our results demonstrate, that even fundamental optical fields such as linearly polarized Gaussian beams can exhibit complex 3D polarization topologies under (tight) focusing conditions. 

\begin{acknowledgments}
We gratefully acknowledge fruitful discussions with Mark R. Dennis and Ebrahim Karimi.
\end{acknowledgments}

\bibliography{bib}

\begin{thebibliography}{41}%
\makeatletter
\providecommand \@ifxundefined [1]{%
 \@ifx{#1\undefined}
}%
\providecommand \@ifnum [1]{%
 \ifnum #1\expandafter \@firstoftwo
 \else \expandafter \@secondoftwo
 \fi
}%
\providecommand \@ifx [1]{%
 \ifx #1\expandafter \@firstoftwo
 \else \expandafter \@secondoftwo
 \fi
}%
\providecommand \natexlab [1]{#1}%
\providecommand \enquote  [1]{``#1''}%
\providecommand \bibnamefont  [1]{#1}%
\providecommand \bibfnamefont [1]{#1}%
\providecommand \citenamefont [1]{#1}%
\providecommand \href@noop [0]{\@secondoftwo}%
\providecommand \href [0]{\begingroup \@sanitize@url \@href}%
\providecommand \@href[1]{\@@startlink{#1}\@@href}%
\providecommand \@@href[1]{\endgroup#1\@@endlink}%
\providecommand \@sanitize@url [0]{\catcode `\\12\catcode `\$12\catcode
  `\&12\catcode `\#12\catcode `\^12\catcode `\_12\catcode `\%12\relax}%
\providecommand \@@startlink[1]{}%
\providecommand \@@endlink[0]{}%
\providecommand \url  [0]{\begingroup\@sanitize@url \@url }%
\providecommand \@url [1]{\endgroup\@href {#1}{\urlprefix }}%
\providecommand \urlprefix  [0]{URL }%
\providecommand \Eprint [0]{\href }%
\providecommand \doibase [0]{http://dx.doi.org/}%
\providecommand \selectlanguage [0]{\@gobble}%
\providecommand \bibinfo  [0]{\@secondoftwo}%
\providecommand \bibfield  [0]{\@secondoftwo}%
\providecommand \translation [1]{[#1]}%
\providecommand \BibitemOpen [0]{}%
\providecommand \bibitemStop [0]{}%
\providecommand \bibitemNoStop [0]{.\EOS\space}%
\providecommand \EOS [0]{\spacefactor3000\relax}%
\providecommand \BibitemShut  [1]{\csname bibitem#1\endcsname}%
\let\auto@bib@innerbib\@empty
\bibitem [{\citenamefont {Quabis}\ \emph {et~al.}(2000)\citenamefont {Quabis},
  \citenamefont {Dorn}, \citenamefont {Eberler}, \citenamefont {Gl\"{o}ckl},\
  and\ \citenamefont {Leuchs}}]{Quabis2000}%
  \BibitemOpen
  \bibfield  {author} {\bibinfo {author} {\bibfnamefont {S.}~\bibnamefont
  {Quabis}}, \bibinfo {author} {\bibfnamefont {R.}~\bibnamefont {Dorn}},
  \bibinfo {author} {\bibfnamefont {M.}~\bibnamefont {Eberler}}, \bibinfo
  {author} {\bibfnamefont {O.}~\bibnamefont {Gl\"{o}ckl}}, \ and\ \bibinfo
  {author} {\bibfnamefont {G.}~\bibnamefont {Leuchs}},\ }\href {\doibase
  10.1016/S0030-4018(99)00729-4} {\bibfield  {journal} {\bibinfo  {journal}
  {Opt. Commun.}\ }\textbf {\bibinfo {volume} {179}},\ \bibinfo {pages} {1}
  (\bibinfo {year} {2000})}\BibitemShut {NoStop}%
\bibitem [{\citenamefont {Youngworth}\ and\ \citenamefont
  {Brown}(2000)}]{Youngworth2000}%
  \BibitemOpen
  \bibfield  {author} {\bibinfo {author} {\bibfnamefont {K.}~\bibnamefont
  {Youngworth}}\ and\ \bibinfo {author} {\bibfnamefont {T.}~\bibnamefont
  {Brown}},\ }\href {\doibase 10.1364/OE.7.000077} {\bibfield  {journal}
  {\bibinfo  {journal} {Opt. Express}\ }\textbf {\bibinfo {volume} {7}},\
  \bibinfo {pages} {77} (\bibinfo {year} {2000})}\BibitemShut {NoStop}%
\bibitem [{\citenamefont {Rittweger}\ \emph {et~al.}(2009)\citenamefont
  {Rittweger}, \citenamefont {Han}, \citenamefont {Irvine}, \citenamefont
  {Eggeling},\ and\ \citenamefont {Hell}}]{Rittweger2009}%
  \BibitemOpen
  \bibfield  {author} {\bibinfo {author} {\bibfnamefont {E.}~\bibnamefont
  {Rittweger}}, \bibinfo {author} {\bibfnamefont {K.~Y.}\ \bibnamefont {Han}},
  \bibinfo {author} {\bibfnamefont {S.~E.}\ \bibnamefont {Irvine}}, \bibinfo
  {author} {\bibfnamefont {C.}~\bibnamefont {Eggeling}}, \ and\ \bibinfo
  {author} {\bibfnamefont {S.~W.}\ \bibnamefont {Hell}},\ }\href {\doibase
  10.1038/nphoton.2009.2} {\bibfield  {journal} {\bibinfo  {journal} {Nature
  Photon.}\ }\textbf {\bibinfo {volume} {3}},\ \bibinfo {pages} {144} (\bibinfo
  {year} {2009})}\BibitemShut {NoStop}%
\bibitem [{\citenamefont {Z\"{u}chner}\ \emph {et~al.}(2008)\citenamefont
  {Z\"{u}chner}, \citenamefont {Failla}, \citenamefont {Hartschuh},\ and\
  \citenamefont {Meixner}}]{Zuchner2008}%
  \BibitemOpen
  \bibfield  {author} {\bibinfo {author} {\bibfnamefont {T.}~\bibnamefont
  {Z\"{u}chner}}, \bibinfo {author} {\bibfnamefont {A.~V.}\ \bibnamefont
  {Failla}}, \bibinfo {author} {\bibfnamefont {A.}~\bibnamefont {Hartschuh}}, \
  and\ \bibinfo {author} {\bibfnamefont {A.~J.}\ \bibnamefont {Meixner}},\
  }\href {\doibase 10.1111/j.1365-2818.2008.01910.x} {\bibfield  {journal}
  {\bibinfo  {journal} {J. Microsc.}\ }\textbf {\bibinfo {volume} {229}},\
  \bibinfo {pages} {337} (\bibinfo {year} {2008})}\BibitemShut {NoStop}%
\bibitem [{\citenamefont {Sancho-Parramon}\ and\ \citenamefont
  {Bosch}(2012)}]{Sancho-Parramon2012}%
  \BibitemOpen
  \bibfield  {author} {\bibinfo {author} {\bibfnamefont {J.}~\bibnamefont
  {Sancho-Parramon}}\ and\ \bibinfo {author} {\bibfnamefont {S.}~\bibnamefont
  {Bosch}},\ }\href {\doibase 10.1021/nn303243p} {\bibfield  {journal}
  {\bibinfo  {journal} {ACS Nano}\ }\textbf {\bibinfo {volume} {6}},\ \bibinfo
  {pages} {8415} (\bibinfo {year} {2012})}\BibitemShut {NoStop}%
\bibitem [{\citenamefont {Neugebauer}\ \emph {et~al.}(2014)\citenamefont
  {Neugebauer}, \citenamefont {Bauer}, \citenamefont {Banzer},\ and\
  \citenamefont {Leuchs}}]{Neugebauer2014}%
  \BibitemOpen
  \bibfield  {author} {\bibinfo {author} {\bibfnamefont {M.}~\bibnamefont
  {Neugebauer}}, \bibinfo {author} {\bibfnamefont {T.}~\bibnamefont {Bauer}},
  \bibinfo {author} {\bibfnamefont {P.}~\bibnamefont {Banzer}}, \ and\ \bibinfo
  {author} {\bibfnamefont {G.}~\bibnamefont {Leuchs}},\ }\href {\doibase
  10.1021/nl5003526} {\bibfield  {journal} {\bibinfo  {journal} {Nano Lett.}\
  }\textbf {\bibinfo {volume} {14}},\ \bibinfo {pages} {2546} (\bibinfo {year}
  {2014})}\BibitemShut {NoStop}%
\bibitem [{\citenamefont {Wo\'{z}niak}\ \emph {et~al.}(2015)\citenamefont
  {Wo\'{z}niak}, \citenamefont {Banzer},\ and\ \citenamefont
  {Leuchs}}]{Wozniak2015}%
  \BibitemOpen
  \bibfield  {author} {\bibinfo {author} {\bibfnamefont {P.}~\bibnamefont
  {Wo\'{z}niak}}, \bibinfo {author} {\bibfnamefont {P.}~\bibnamefont {Banzer}},
  \ and\ \bibinfo {author} {\bibfnamefont {G.}~\bibnamefont {Leuchs}},\ }\href
  {\doibase 10.1002/lpor.201400188} {\bibfield  {journal} {\bibinfo  {journal}
  {Laser Photon. Rev.}\ }\textbf {\bibinfo {volume} {9}},\ \bibinfo {pages}
  {231} (\bibinfo {year} {2015})}\BibitemShut {NoStop}%
\bibitem [{\citenamefont {Dholakia}\ and\ \citenamefont
  {{\v{C}}i{\v{z}}m{\'{a}}r}(2011)}]{Dholakia2011}%
  \BibitemOpen
  \bibfield  {author} {\bibinfo {author} {\bibfnamefont {K.}~\bibnamefont
  {Dholakia}}\ and\ \bibinfo {author} {\bibfnamefont {T.}~\bibnamefont
  {{\v{C}}i{\v{z}}m{\'{a}}r}},\ }\href {\doibase 10.1038/nphoton.2011.80}
  {\bibfield  {journal} {\bibinfo  {journal} {Nature Photon.}\ }\textbf
  {\bibinfo {volume} {5}},\ \bibinfo {pages} {335} (\bibinfo {year}
  {2011})}\BibitemShut {NoStop}%
\bibitem [{\citenamefont {Taylor}\ \emph {et~al.}(2015)\citenamefont {Taylor},
  \citenamefont {Waleed}, \citenamefont {Stilgoe}, \citenamefont
  {Rubinsztein-Dunlop},\ and\ \citenamefont {Bowen}}]{Taylor2015}%
  \BibitemOpen
  \bibfield  {author} {\bibinfo {author} {\bibfnamefont {M.~A.}\ \bibnamefont
  {Taylor}}, \bibinfo {author} {\bibfnamefont {M.}~\bibnamefont {Waleed}},
  \bibinfo {author} {\bibfnamefont {A.~B.}\ \bibnamefont {Stilgoe}}, \bibinfo
  {author} {\bibfnamefont {H.}~\bibnamefont {Rubinsztein-Dunlop}}, \ and\
  \bibinfo {author} {\bibfnamefont {W.~P.}\ \bibnamefont {Bowen}},\ }\href
  {\doibase 10.1038/nphoton.2015.160} {\bibfield  {journal} {\bibinfo
  {journal} {Nature Photon.}\ }\textbf {\bibinfo {volume} {9}},\ \bibinfo
  {pages} {669} (\bibinfo {year} {2015})}\BibitemShut {NoStop}%
\bibitem [{\citenamefont {Nye}\ and\ \citenamefont {Berry}(1974)}]{Nye1974}%
  \BibitemOpen
  \bibfield  {author} {\bibinfo {author} {\bibfnamefont {J.~F.}\ \bibnamefont
  {Nye}}\ and\ \bibinfo {author} {\bibfnamefont {M.~V.}\ \bibnamefont
  {Berry}},\ }\href {\doibase 10.1098/rspa.1974.0012} {\bibfield  {journal}
  {\bibinfo  {journal} {Proc. R. Soc. A}\ }\textbf {\bibinfo {volume} {336}},\
  \bibinfo {pages} {165} (\bibinfo {year} {1974})}\BibitemShut {NoStop}%
\bibitem [{\citenamefont {Allen}\ \emph {et~al.}(1992)\citenamefont {Allen},
  \citenamefont {Beijersbergen}, \citenamefont {Spreeuw},\ and\ \citenamefont
  {Woerdman}}]{Allen1992}%
  \BibitemOpen
  \bibfield  {author} {\bibinfo {author} {\bibfnamefont {L.}~\bibnamefont
  {Allen}}, \bibinfo {author} {\bibfnamefont {M.~W.}\ \bibnamefont
  {Beijersbergen}}, \bibinfo {author} {\bibfnamefont {R.~J.~C.}\ \bibnamefont
  {Spreeuw}}, \ and\ \bibinfo {author} {\bibfnamefont {J.~P.}\ \bibnamefont
  {Woerdman}},\ }\href {\doibase 10.1103/PhysRevA.45.8185} {\bibfield
  {journal} {\bibinfo  {journal} {Phys. Rev. A}\ }\textbf {\bibinfo {volume}
  {45}},\ \bibinfo {pages} {8185} (\bibinfo {year} {1992})}\BibitemShut
  {NoStop}%
\bibitem [{\citenamefont {Leach}\ \emph {et~al.}(2004)\citenamefont {Leach},
  \citenamefont {Dennis}, \citenamefont {Courtial},\ and\ \citenamefont
  {Padgett}}]{Leach2004}%
  \BibitemOpen
  \bibfield  {author} {\bibinfo {author} {\bibfnamefont {J.}~\bibnamefont
  {Leach}}, \bibinfo {author} {\bibfnamefont {M.~R.}\ \bibnamefont {Dennis}},
  \bibinfo {author} {\bibfnamefont {J.}~\bibnamefont {Courtial}}, \ and\
  \bibinfo {author} {\bibfnamefont {M.~J.}\ \bibnamefont {Padgett}},\ }\href
  {\doibase 10.1038/nature03092} {\bibfield  {journal} {\bibinfo  {journal}
  {Nature}\ }\textbf {\bibinfo {volume} {432}},\ \bibinfo {pages} {165}
  (\bibinfo {year} {2004})}\BibitemShut {NoStop}%
\bibitem [{\citenamefont {Dennis}\ \emph {et~al.}(2010)\citenamefont {Dennis},
  \citenamefont {King}, \citenamefont {Jack}, \citenamefont {O'Holleran},\ and\
  \citenamefont {Padgett}}]{Dennis2010}%
  \BibitemOpen
  \bibfield  {author} {\bibinfo {author} {\bibfnamefont {M.~R.}\ \bibnamefont
  {Dennis}}, \bibinfo {author} {\bibfnamefont {R.~P.}\ \bibnamefont {King}},
  \bibinfo {author} {\bibfnamefont {B.}~\bibnamefont {Jack}}, \bibinfo {author}
  {\bibfnamefont {K.}~\bibnamefont {O'Holleran}}, \ and\ \bibinfo {author}
  {\bibfnamefont {M.~J.}\ \bibnamefont {Padgett}},\ }\href {\doibase
  10.1038/nphys1504} {\bibfield  {journal} {\bibinfo  {journal} {Nature Phys.}\
  }\textbf {\bibinfo {volume} {6}},\ \bibinfo {pages} {118} (\bibinfo {year}
  {2010})}\BibitemShut {NoStop}%
\bibitem [{\citenamefont {Freund}(2005)}]{Freund2005}%
  \BibitemOpen
  \bibfield  {author} {\bibinfo {author} {\bibfnamefont {I.}~\bibnamefont
  {Freund}},\ }\href {\doibase 10.1016/j.optcom.2004.12.052} {\bibfield
  {journal} {\bibinfo  {journal} {Opt. Commun.}\ }\textbf {\bibinfo {volume}
  {249}},\ \bibinfo {pages} {7} (\bibinfo {year} {2005})}\BibitemShut {NoStop}%
\bibitem [{\citenamefont {Dennis}(2011)}]{Dennis2011}%
  \BibitemOpen
  \bibfield  {author} {\bibinfo {author} {\bibfnamefont {M.~R.}\ \bibnamefont
  {Dennis}},\ }\href {\doibase 10.1364/OL.36.003765} {\bibfield  {journal}
  {\bibinfo  {journal} {Opt. Lett.}\ }\textbf {\bibinfo {volume} {36}},\
  \bibinfo {pages} {3765} (\bibinfo {year} {2011})}\BibitemShut {NoStop}%
\bibitem [{\citenamefont {Bauer}\ \emph {et~al.}(2015)\citenamefont {Bauer},
  \citenamefont {Banzer}, \citenamefont {Karimi}, \citenamefont {Orlov},
  \citenamefont {Rubano}, \citenamefont {Marrucci}, \citenamefont {Santamato},
  \citenamefont {Boyd},\ and\ \citenamefont {Leuchs}}]{Bauer2015}%
  \BibitemOpen
  \bibfield  {author} {\bibinfo {author} {\bibfnamefont {T.}~\bibnamefont
  {Bauer}}, \bibinfo {author} {\bibfnamefont {P.}~\bibnamefont {Banzer}},
  \bibinfo {author} {\bibfnamefont {E.}~\bibnamefont {Karimi}}, \bibinfo
  {author} {\bibfnamefont {S.}~\bibnamefont {Orlov}}, \bibinfo {author}
  {\bibfnamefont {A.}~\bibnamefont {Rubano}}, \bibinfo {author} {\bibfnamefont
  {L.}~\bibnamefont {Marrucci}}, \bibinfo {author} {\bibfnamefont
  {E.}~\bibnamefont {Santamato}}, \bibinfo {author} {\bibfnamefont {R.~W.}\
  \bibnamefont {Boyd}}, \ and\ \bibinfo {author} {\bibfnamefont
  {G.}~\bibnamefont {Leuchs}},\ }\href {\doibase 10.1126/science.1260635}
  {\bibfield  {journal} {\bibinfo  {journal} {Science}\ }\textbf {\bibinfo
  {volume} {347}},\ \bibinfo {pages} {27} (\bibinfo {year} {2015})}\BibitemShut
  {NoStop}%
\bibitem [{\citenamefont {Nye}\ and\ \citenamefont {Hajnal}(1987)}]{Nye1987}%
  \BibitemOpen
  \bibfield  {author} {\bibinfo {author} {\bibfnamefont {J.~F.}\ \bibnamefont
  {Nye}}\ and\ \bibinfo {author} {\bibfnamefont {J.~V.}\ \bibnamefont
  {Hajnal}},\ }\href {\doibase 10.1098/rspa.1987.0002} {\bibfield  {journal}
  {\bibinfo  {journal} {Proc. R. Soc. A}\ }\textbf {\bibinfo {volume} {409}},\
  \bibinfo {pages} {21} (\bibinfo {year} {1987})}\BibitemShut {NoStop}%
\bibitem [{\citenamefont {Freund}(2010)}]{Freund2010}%
  \BibitemOpen
  \bibfield  {author} {\bibinfo {author} {\bibfnamefont {I.}~\bibnamefont
  {Freund}},\ }\href {\doibase 10.1364/OL.35.000148} {\bibfield  {journal}
  {\bibinfo  {journal} {Opt. Lett.}\ }\textbf {\bibinfo {volume} {35}},\
  \bibinfo {pages} {148} (\bibinfo {year} {2010})}\BibitemShut {NoStop}%
\bibitem [{\citenamefont {Bauer}\ \emph {et~al.}(2014)\citenamefont {Bauer},
  \citenamefont {Orlov}, \citenamefont {Peschel}, \citenamefont {Banzer},\ and\
  \citenamefont {Leuchs}}]{Bauer2013}%
  \BibitemOpen
  \bibfield  {author} {\bibinfo {author} {\bibfnamefont {T.}~\bibnamefont
  {Bauer}}, \bibinfo {author} {\bibfnamefont {S.}~\bibnamefont {Orlov}},
  \bibinfo {author} {\bibfnamefont {U.}~\bibnamefont {Peschel}}, \bibinfo
  {author} {\bibfnamefont {P.}~\bibnamefont {Banzer}}, \ and\ \bibinfo {author}
  {\bibfnamefont {G.}~\bibnamefont {Leuchs}},\ }\href {\doibase
  10.1038/nphoton.2013.289} {\bibfield  {journal} {\bibinfo  {journal} {Nature
  Photon.}\ }\textbf {\bibinfo {volume} {8}},\ \bibinfo {pages} {23} (\bibinfo
  {year} {2014})}\BibitemShut {NoStop}%
\bibitem [{\citenamefont {Beckley}\ \emph {et~al.}(2010)\citenamefont
  {Beckley}, \citenamefont {Brown},\ and\ \citenamefont
  {Alonso}}]{Beckley2010}%
  \BibitemOpen
  \bibfield  {author} {\bibinfo {author} {\bibfnamefont {A.~M.}\ \bibnamefont
  {Beckley}}, \bibinfo {author} {\bibfnamefont {T.~G.}\ \bibnamefont {Brown}},
  \ and\ \bibinfo {author} {\bibfnamefont {M.~A.}\ \bibnamefont {Alonso}},\
  }\href {\doibase 10.1364/OE.18.010777} {\bibfield  {journal} {\bibinfo
  {journal} {Opt. Express}\ }\textbf {\bibinfo {volume} {18}},\ \bibinfo
  {pages} {10777} (\bibinfo {year} {2010})}\BibitemShut {NoStop}%
\bibitem [{\citenamefont {Galvez}\ \emph {et~al.}(2012)\citenamefont {Galvez},
  \citenamefont {Khadka}, \citenamefont {Schubert},\ and\ \citenamefont
  {Nomoto}}]{Galvez2012}%
  \BibitemOpen
  \bibfield  {author} {\bibinfo {author} {\bibfnamefont {E.~J.}\ \bibnamefont
  {Galvez}}, \bibinfo {author} {\bibfnamefont {S.}~\bibnamefont {Khadka}},
  \bibinfo {author} {\bibfnamefont {W.~H.}\ \bibnamefont {Schubert}}, \ and\
  \bibinfo {author} {\bibfnamefont {S.}~\bibnamefont {Nomoto}},\ }\href
  {\doibase 10.1364/AO.51.002925} {\bibfield  {journal} {\bibinfo  {journal}
  {Appl. Opt.}\ }\textbf {\bibinfo {volume} {51}},\ \bibinfo {pages} {2925 }
  (\bibinfo {year} {2012})}\BibitemShut {NoStop}%
\bibitem [{\citenamefont {Cardano}\ \emph {et~al.}(2013)\citenamefont
  {Cardano}, \citenamefont {Karimi}, \citenamefont {Marrucci}, \citenamefont
  {de~Lisio},\ and\ \citenamefont {Santamato}}]{Cardano2013}%
  \BibitemOpen
  \bibfield  {author} {\bibinfo {author} {\bibfnamefont {F.}~\bibnamefont
  {Cardano}}, \bibinfo {author} {\bibfnamefont {E.}~\bibnamefont {Karimi}},
  \bibinfo {author} {\bibfnamefont {L.}~\bibnamefont {Marrucci}}, \bibinfo
  {author} {\bibfnamefont {C.}~\bibnamefont {de~Lisio}}, \ and\ \bibinfo
  {author} {\bibfnamefont {E.}~\bibnamefont {Santamato}},\ }\href {\doibase
  10.1364/OE.21.008815} {\bibfield  {journal} {\bibinfo  {journal} {Opt.
  Express}\ }\textbf {\bibinfo {volume} {21}},\ \bibinfo {pages} {8815}
  (\bibinfo {year} {2013})}\BibitemShut {NoStop}%
\bibitem [{\citenamefont {Aiello}\ \emph {et~al.}(2015)\citenamefont {Aiello},
  \citenamefont {Banzer}, \citenamefont {Neugebauer},\ and\ \citenamefont
  {Leuchs}}]{Aiello2015-2}%
  \BibitemOpen
  \bibfield  {author} {\bibinfo {author} {\bibfnamefont {A.}~\bibnamefont
  {Aiello}}, \bibinfo {author} {\bibfnamefont {P.}~\bibnamefont {Banzer}},
  \bibinfo {author} {\bibfnamefont {M.}~\bibnamefont {Neugebauer}}, \ and\
  \bibinfo {author} {\bibfnamefont {G.}~\bibnamefont {Leuchs}},\ }\href
  {\doibase 10.1038/nphoton.2015.203} {\bibfield  {journal} {\bibinfo
  {journal} {Nature Photonics}\ }\textbf {\bibinfo {volume} {9}},\ \bibinfo
  {pages} {789} (\bibinfo {year} {2015})}\BibitemShut {NoStop}%
\bibitem [{\citenamefont {Bliokh}\ and\ \citenamefont
  {Nori}(2012)}]{Bliokh2012}%
  \BibitemOpen
  \bibfield  {author} {\bibinfo {author} {\bibfnamefont {K.~Y.}\ \bibnamefont
  {Bliokh}}\ and\ \bibinfo {author} {\bibfnamefont {F.}~\bibnamefont {Nori}},\
  }\href {\doibase 10.1103/PhysRevA.85.061801} {\bibfield  {journal} {\bibinfo
  {journal} {Phys. Rev. A}\ }\textbf {\bibinfo {volume} {85}},\ \bibinfo
  {pages} {061801} (\bibinfo {year} {2012})}\BibitemShut {NoStop}%
\bibitem [{\citenamefont {Rodr{\'{\i}}guez-Fortu{\~{n}}o}\ \emph
  {et~al.}(2013)\citenamefont {Rodr{\'{\i}}guez-Fortu{\~{n}}o}, \citenamefont
  {Marino}, \citenamefont {Ginzburg}, \citenamefont {O'Connor}, \citenamefont
  {Mart{\'{\i}}nez}, \citenamefont {Wurtz},\ and\ \citenamefont
  {Zayats}}]{Rodriguez-Fortuno2013}%
  \BibitemOpen
  \bibfield  {author} {\bibinfo {author} {\bibfnamefont {F.~J.}\ \bibnamefont
  {Rodr{\'{\i}}guez-Fortu{\~{n}}o}}, \bibinfo {author} {\bibfnamefont
  {G.}~\bibnamefont {Marino}}, \bibinfo {author} {\bibfnamefont
  {P.}~\bibnamefont {Ginzburg}}, \bibinfo {author} {\bibfnamefont
  {D.}~\bibnamefont {O'Connor}}, \bibinfo {author} {\bibfnamefont
  {A.}~\bibnamefont {Mart{\'{\i}}nez}}, \bibinfo {author} {\bibfnamefont
  {G.~A.}\ \bibnamefont {Wurtz}}, \ and\ \bibinfo {author} {\bibfnamefont
  {A.~V.}\ \bibnamefont {Zayats}},\ }\href {\doibase 10.1126/science.1233739}
  {\bibfield  {journal} {\bibinfo  {journal} {Science}\ }\textbf {\bibinfo
  {volume} {340}},\ \bibinfo {pages} {328} (\bibinfo {year}
  {2013})}\BibitemShut {NoStop}%
\bibitem [{\citenamefont {Petersen}\ \emph {et~al.}(2014)\citenamefont
  {Petersen}, \citenamefont {Volz},\ and\ \citenamefont
  {Rauschenbeutel}}]{Petersen2014}%
  \BibitemOpen
  \bibfield  {author} {\bibinfo {author} {\bibfnamefont {J.}~\bibnamefont
  {Petersen}}, \bibinfo {author} {\bibfnamefont {J.}~\bibnamefont {Volz}}, \
  and\ \bibinfo {author} {\bibfnamefont {A.}~\bibnamefont {Rauschenbeutel}},\
  }\href {\doibase 10.1126/science.1257671} {\bibfield  {journal} {\bibinfo
  {journal} {Science}\ }\textbf {\bibinfo {volume} {346}},\ \bibinfo {pages}
  {67} (\bibinfo {year} {2014})}\BibitemShut {NoStop}%
\bibitem [{\citenamefont {Yang}\ and\ \citenamefont {Cohen}(2011)}]{Yang2011}%
  \BibitemOpen
  \bibfield  {author} {\bibinfo {author} {\bibfnamefont {N.}~\bibnamefont
  {Yang}}\ and\ \bibinfo {author} {\bibfnamefont {A.~E.}\ \bibnamefont
  {Cohen}},\ }\href {\doibase 10.1021/jp1092898} {\bibfield  {journal}
  {\bibinfo  {journal} {J. Phys. Chem. B}\ }\textbf {\bibinfo {volume} {115}},\
  \bibinfo {pages} {5304} (\bibinfo {year} {2011})}\BibitemShut {NoStop}%
\bibitem [{\citenamefont {Banzer}\ \emph {et~al.}(2013)\citenamefont {Banzer},
  \citenamefont {Neugebauer}, \citenamefont {Aiello}, \citenamefont
  {Marquardt}, \citenamefont {Lindlein}, \citenamefont {Bauer},\ and\
  \citenamefont {Leuchs}}]{Banzer2013}%
  \BibitemOpen
  \bibfield  {author} {\bibinfo {author} {\bibfnamefont {P.}~\bibnamefont
  {Banzer}}, \bibinfo {author} {\bibfnamefont {M.}~\bibnamefont {Neugebauer}},
  \bibinfo {author} {\bibfnamefont {A.}~\bibnamefont {Aiello}}, \bibinfo
  {author} {\bibfnamefont {C.}~\bibnamefont {Marquardt}}, \bibinfo {author}
  {\bibfnamefont {N.}~\bibnamefont {Lindlein}}, \bibinfo {author}
  {\bibfnamefont {T.}~\bibnamefont {Bauer}}, \ and\ \bibinfo {author}
  {\bibfnamefont {G.}~\bibnamefont {Leuchs}},\ }\href {\doibase
  10.2971/jeos.2013.13032} {\bibfield  {journal} {\bibinfo  {journal} {J. Eur.
  Opt. Soc, Rapid Publ.}\ }\textbf {\bibinfo {volume} {8}},\ \bibinfo {pages}
  {13032} (\bibinfo {year} {2013})}\BibitemShut {NoStop}%
\bibitem [{\citenamefont {Neugebauer}\ \emph {et~al.}(2015)\citenamefont
  {Neugebauer}, \citenamefont {Bauer}, \citenamefont {Aiello},\ and\
  \citenamefont {Banzer}}]{Neugebauer2015}%
  \BibitemOpen
  \bibfield  {author} {\bibinfo {author} {\bibfnamefont {M.}~\bibnamefont
  {Neugebauer}}, \bibinfo {author} {\bibfnamefont {T.}~\bibnamefont {Bauer}},
  \bibinfo {author} {\bibfnamefont {A.}~\bibnamefont {Aiello}}, \ and\ \bibinfo
  {author} {\bibfnamefont {P.}~\bibnamefont {Banzer}},\ }\href {\doibase
  10.1103/PhysRevLett.114.063901} {\bibfield  {journal} {\bibinfo  {journal}
  {Phys. Rev. Lett.}\ }\textbf {\bibinfo {volume} {114}},\ \bibinfo {pages}
  {063901} (\bibinfo {year} {2015})}\BibitemShut {NoStop}%
\bibitem [{\citenamefont {Richards}\ and\ \citenamefont
  {Wolf}(1959)}]{Richards1959}%
  \BibitemOpen
  \bibfield  {author} {\bibinfo {author} {\bibfnamefont {B.}~\bibnamefont
  {Richards}}\ and\ \bibinfo {author} {\bibfnamefont {E.}~\bibnamefont
  {Wolf}},\ }\href {\doibase 10.1098/rspa.1959.0200} {\bibfield  {journal}
  {\bibinfo  {journal} {Proc. R. Soc. A}\ }\textbf {\bibinfo {volume} {253}},\
  \bibinfo {pages} {358} (\bibinfo {year} {1959})}\BibitemShut {NoStop}%
\bibitem [{\citenamefont {Berry}(2004)}]{Berry2004}%
  \BibitemOpen
  \bibfield  {author} {\bibinfo {author} {\bibfnamefont {M.~V.}\ \bibnamefont
  {Berry}},\ }\href {\doibase 10.1088/1464-4258/6/7/003} {\bibfield  {journal}
  {\bibinfo  {journal} {J. Opt. A}\ }\textbf {\bibinfo {volume} {6}},\ \bibinfo
  {pages} {675} (\bibinfo {year} {2004})}\BibitemShut {NoStop}%
\bibitem [{\citenamefont {Berry}\ and\ \citenamefont
  {Dennis}(2001)}]{Berry2001}%
  \BibitemOpen
  \bibfield  {author} {\bibinfo {author} {\bibfnamefont {M.~V.}\ \bibnamefont
  {Berry}}\ and\ \bibinfo {author} {\bibfnamefont {M.~R.}\ \bibnamefont
  {Dennis}},\ }\href {\doibase 10.1098/rspa.2000.0660} {\bibfield  {journal}
  {\bibinfo  {journal} {Proc. R. Soc. A}\ }\textbf {\bibinfo {volume} {457}},\
  \bibinfo {pages} {141} (\bibinfo {year} {2001})}\BibitemShut {NoStop}%
\bibitem [{\citenamefont {Bliokh}\ \emph {et~al.}(2014)\citenamefont {Bliokh},
  \citenamefont {Bekshaev},\ and\ \citenamefont {Nori}}]{Bliokh2014}%
  \BibitemOpen
  \bibfield  {author} {\bibinfo {author} {\bibfnamefont {K.~Y.}\ \bibnamefont
  {Bliokh}}, \bibinfo {author} {\bibfnamefont {A.~Y.}\ \bibnamefont
  {Bekshaev}}, \ and\ \bibinfo {author} {\bibfnamefont {F.}~\bibnamefont
  {Nori}},\ }\href {\doibase 10.1038/ncomms4300} {\bibfield  {journal}
  {\bibinfo  {journal} {Nat. Commun.}\ }\textbf {\bibinfo {volume} {5}},\
  \bibinfo {pages} {3300} (\bibinfo {year} {2014})}\BibitemShut {NoStop}%
\bibitem [{\citenamefont {Aiello}\ and\ \citenamefont
  {Banzer}(2015)}]{Aiello2015}%
  \BibitemOpen
  \bibfield  {author} {\bibinfo {author} {\bibfnamefont {A.}~\bibnamefont
  {Aiello}}\ and\ \bibinfo {author} {\bibfnamefont {P.}~\bibnamefont
  {Banzer}},\ }\href {http://arxiv.org/abs/1502.05350} {\bibfield  {journal}
  {\bibinfo  {journal} {Arxiv preprint arXiv:1502.05350}\ } (\bibinfo {year}
  {2015})}\BibitemShut {NoStop}%
\bibitem [{\citenamefont {Berry}\ and\ \citenamefont
  {Hannay}(1977)}]{Berry1977}%
  \BibitemOpen
  \bibfield  {author} {\bibinfo {author} {\bibfnamefont {M.~V.}\ \bibnamefont
  {Berry}}\ and\ \bibinfo {author} {\bibfnamefont {J.~H.}\ \bibnamefont
  {Hannay}},\ }\href {\doibase 10.1088/0305-4470/10/11/009} {\bibfield
  {journal} {\bibinfo  {journal} {J. Phys. A}\ }\textbf {\bibinfo {volume}
  {10}},\ \bibinfo {pages} {1809} (\bibinfo {year} {1977})}\BibitemShut
  {NoStop}%
\bibitem [{\citenamefont {Novotny}\ and\ \citenamefont
  {Hecht}(2006)}]{L.NovotnyandB.Hecht2006}%
  \BibitemOpen
  \bibfield  {author} {\bibinfo {author} {\bibfnamefont {L.}~\bibnamefont
  {Novotny}}\ and\ \bibinfo {author} {\bibfnamefont {B.}~\bibnamefont
  {Hecht}},\ }\href@noop {} {\emph {\bibinfo {title} {{Principles of
  Nano-Optics}}}}\ (\bibinfo  {publisher} {Cambridge University Press},\
  \bibinfo {address} {Cambridge},\ \bibinfo {year} {2006})\BibitemShut
  {NoStop}%
\bibitem [{\citenamefont {Dorn}\ \emph {et~al.}(2003)\citenamefont {Dorn},
  \citenamefont {Quabis},\ and\ \citenamefont {Leuchs}}]{Dorn2003}%
  \BibitemOpen
  \bibfield  {author} {\bibinfo {author} {\bibfnamefont {R.}~\bibnamefont
  {Dorn}}, \bibinfo {author} {\bibfnamefont {S.}~\bibnamefont {Quabis}}, \ and\
  \bibinfo {author} {\bibfnamefont {G.}~\bibnamefont {Leuchs}},\ }\href
  {\doibase 10.1080/09500340308235246} {\bibfield  {journal} {\bibinfo
  {journal} {J. Mod. Opt.}\ }\textbf {\bibinfo {volume} {50}},\ \bibinfo
  {pages} {1917} (\bibinfo {year} {2003})}\BibitemShut {NoStop}%
\bibitem [{\citenamefont {Novotny}\ \emph {et~al.}(2001)\citenamefont
  {Novotny}, \citenamefont {Grober},\ and\ \citenamefont
  {Karrai}}]{Novotny2001}%
  \BibitemOpen
  \bibfield  {author} {\bibinfo {author} {\bibfnamefont {L.}~\bibnamefont
  {Novotny}}, \bibinfo {author} {\bibfnamefont {R.~D.}\ \bibnamefont {Grober}},
  \ and\ \bibinfo {author} {\bibfnamefont {K.}~\bibnamefont {Karrai}},\ }\href
  {\doibase 10.1364/OL.26.000789} {\bibfield  {journal} {\bibinfo  {journal}
  {Opt. Lett.}\ }\textbf {\bibinfo {volume} {26}},\ \bibinfo {pages} {789}
  (\bibinfo {year} {2001})}\BibitemShut {NoStop}%
\bibitem [{\citenamefont {Bahlmann}\ and\ \citenamefont
  {Hell}(2000)}]{Bahlmann2000}%
  \BibitemOpen
  \bibfield  {author} {\bibinfo {author} {\bibfnamefont {K.}~\bibnamefont
  {Bahlmann}}\ and\ \bibinfo {author} {\bibfnamefont {S.~W.}\ \bibnamefont
  {Hell}},\ }\href {\doibase 10.1063/1.127061} {\bibfield  {journal} {\bibinfo
  {journal} {Appl. Phys. Lett.}\ }\textbf {\bibinfo {volume} {77}},\ \bibinfo
  {pages} {612} (\bibinfo {year} {2000})}\BibitemShut {NoStop}%
\bibitem [{\citenamefont {Banzer}\ \emph {et~al.}(2010)\citenamefont {Banzer},
  \citenamefont {Peschel}, \citenamefont {Quabis},\ and\ \citenamefont
  {Leuchs}}]{Banzer2010a}%
  \BibitemOpen
  \bibfield  {author} {\bibinfo {author} {\bibfnamefont {P.}~\bibnamefont
  {Banzer}}, \bibinfo {author} {\bibfnamefont {U.}~\bibnamefont {Peschel}},
  \bibinfo {author} {\bibfnamefont {S.}~\bibnamefont {Quabis}}, \ and\ \bibinfo
  {author} {\bibfnamefont {G.}~\bibnamefont {Leuchs}},\ }\href {\doibase
  10.1364/OE.18.010905} {\bibfield  {journal} {\bibinfo  {journal} {Opt.
  Express}\ }\textbf {\bibinfo {volume} {18}},\ \bibinfo {pages} {10905}
  (\bibinfo {year} {2010})}\BibitemShut {NoStop}%
\bibitem [{Note1()}]{Note1}%
  \BibitemOpen
  \bibinfo {note} {See Supplemental Material at [URL will be inserted by
  publisher] for details about the experimentally measured interferometric data
  and reconstructed focal field distribution.}\BibitemShut {Stop}%
\end{thebibliography}%


\begin{thebibliography}{3}%
\makeatletter
\providecommand \@ifxundefined [1]{%
 \@ifx{#1\undefined}
}%
\providecommand \@ifnum [1]{%
 \ifnum #1\expandafter \@firstoftwo
 \else \expandafter \@secondoftwo
 \fi
}%
\providecommand \@ifx [1]{%
 \ifx #1\expandafter \@firstoftwo
 \else \expandafter \@secondoftwo
 \fi
}%
\providecommand \natexlab [1]{#1}%
\providecommand \enquote  [1]{``#1''}%
\providecommand \bibnamefont  [1]{#1}%
\providecommand \bibfnamefont [1]{#1}%
\providecommand \citenamefont [1]{#1}%
\providecommand \href@noop [0]{\@secondoftwo}%
\providecommand \href [0]{\begingroup \@sanitize@url \@href}%
\providecommand \@href[1]{\@@startlink{#1}\@@href}%
\providecommand \@@href[1]{\endgroup#1\@@endlink}%
\providecommand \@sanitize@url [0]{\catcode `\\12\catcode `\$12\catcode
  `\&12\catcode `\#12\catcode `\^12\catcode `\_12\catcode `\%12\relax}%
\providecommand \@@startlink[1]{}%
\providecommand \@@endlink[0]{}%
\providecommand \url  [0]{\begingroup\@sanitize@url \@url }%
\providecommand \@url [1]{\endgroup\@href {#1}{\urlprefix }}%
\providecommand \urlprefix  [0]{URL }%
\providecommand \Eprint [0]{\href }%
\providecommand \doibase [0]{http://dx.doi.org/}%
\providecommand \selectlanguage [0]{\@gobble}%
\providecommand \bibinfo  [0]{\@secondoftwo}%
\providecommand \bibfield  [0]{\@secondoftwo}%
\providecommand \translation [1]{[#1]}%
\providecommand \BibitemOpen [0]{}%
\providecommand \bibitemStop [0]{}%
\providecommand \bibitemNoStop [0]{.\EOS\space}%
\providecommand \EOS [0]{\spacefactor3000\relax}%
\providecommand \BibitemShut  [1]{\csname bibitem#1\endcsname}%
\let\auto@bib@innerbib\@empty
\bibitem [{\citenamefont {Bauer}\ \emph {et~al.}(2014)\citenamefont {Bauer},
  \citenamefont {Orlov}, \citenamefont {Peschel}, \citenamefont {Banzer},\ and\
  \citenamefont {Leuchs}}]{Bauer2013}%
  \BibitemOpen
  \bibfield  {author} {\bibinfo {author} {\bibfnamefont {T.}~\bibnamefont
  {Bauer}}, \bibinfo {author} {\bibfnamefont {S.}~\bibnamefont {Orlov}},
  \bibinfo {author} {\bibfnamefont {U.}~\bibnamefont {Peschel}}, \bibinfo
  {author} {\bibfnamefont {P.}~\bibnamefont {Banzer}}, \ and\ \bibinfo {author}
  {\bibfnamefont {G.}~\bibnamefont {Leuchs}},\ }\href {\doibase
  10.1038/nphoton.2013.289} {\bibfield  {journal} {\bibinfo  {journal} {Nature
  Photon.}\ }\textbf {\bibinfo {volume} {8}},\ \bibinfo {pages} {23} (\bibinfo
  {year} {2014})}\BibitemShut {NoStop}%
\bibitem [{\citenamefont {Tsang}\ \emph {et~al.}(2000)\citenamefont {Tsang},
  \citenamefont {Kong},\ and\ \citenamefont {Ding}}]{Tsang2000}%
  \BibitemOpen
  \bibfield  {author} {\bibinfo {author} {\bibfnamefont {L.}~\bibnamefont
  {Tsang}}, \bibinfo {author} {\bibfnamefont {J.~A.}\ \bibnamefont {Kong}}, \
  and\ \bibinfo {author} {\bibfnamefont {K.-H.}\ \bibnamefont {Ding}},\
  }\href@noop {} {\emph {\bibinfo {title} {{Scattering of electromagnetic
  waves}}}},\ \bibinfo {edition} {1st}\ ed.\ (\bibinfo  {publisher} {John Wiley
  {\&} Sons, Inc.},\ \bibinfo {address} {New York},\ \bibinfo {year}
  {2000})\BibitemShut {NoStop}%
\bibitem [{\citenamefont {Novotny}\ and\ \citenamefont
  {Hecht}(2006)}]{L.NovotnyandB.Hecht2006}%
  \BibitemOpen
  \bibfield  {author} {\bibinfo {author} {\bibfnamefont {L.}~\bibnamefont
  {Novotny}}\ and\ \bibinfo {author} {\bibfnamefont {B.}~\bibnamefont
  {Hecht}},\ }\href@noop {} {\emph {\bibinfo {title} {{Principles of
  Nano-Optics}}}}\ (\bibinfo  {publisher} {Cambridge University Press},\
  \bibinfo {address} {Cambridge},\ \bibinfo {year} {2006})\BibitemShut
  {NoStop}%
\end{thebibliography}%

\end{document}


\title{Supplementary Materials to: Optical Polarization M\"obius Strips and Points of Purely Transverse Spin Density}

\author{Thomas Bauer}
\affiliation{Max Planck Institute for the Science of Light, Guenther-Scharowsky-Str. 1, D-91058 Erlangen, Germany}
\affiliation{Institute of Optics, Information and Photonics, University Erlangen-Nuremberg, Staudtstr. 7/B2, D-91058 Erlangen, Germany}
\author{Martin Neugebauer}
\affiliation{Max Planck Institute for the Science of Light, Guenther-Scharowsky-Str. 1, D-91058 Erlangen, Germany}
\affiliation{Institute of Optics, Information and Photonics, University Erlangen-Nuremberg, Staudtstr. 7/B2, D-91058 Erlangen, Germany}
\author{Gerd Leuchs}
\affiliation{Max Planck Institute for the Science of Light, Guenther-Scharowsky-Str. 1, D-91058 Erlangen, Germany}
\affiliation{Institute of Optics, Information and Photonics, University Erlangen-Nuremberg, Staudtstr. 7/B2, D-91058 Erlangen, Germany}
\affiliation{Department of Physics, University of Ottawa, 25 Templeton, Ottawa, Ontario K1N 6N5 Canada}
\author{Peter Banzer}
\email[]{peter.banzer@mpl.mpg.de}
\homepage[]{http://www.mpl.mpg.de/}
\affiliation{Max Planck Institute for the Science of Light, Guenther-Scharowsky-Str. 1, D-91058 Erlangen, Germany}
\affiliation{Institute of Optics, Information and Photonics, University Erlangen-Nuremberg, Staudtstr. 7/B2, D-91058 Erlangen, Germany}
\affiliation{Department of Physics, University of Ottawa, 25 Templeton, Ottawa, Ontario K1N 6N5 Canada}

\date{\today}

\begin{abstract}
In these supplementary materials, we show the detailed experimentally measured data of a linearly $y$-polarized tightly focused Gaussian beam and the corresponding focal field distribution, resulting from a nanointerferometric reconstruction technique \cite{Bauer2013}. A short overview of the employed scheme is given and the spin density distributions derived from the experimentally reconstructed focal field are shown. To highlight the experimental occurrence of polarization M\"obius strips in tightly focused light fields with minute phase aberrations, even around C-points with purely transverse spin, the experimentally realized trace of the major axis of the polarization ellipse is shown around such a C-point with purely transverse spin.
\end{abstract}

\pacs{03.50.De, 42.25.Ja, 42.50.Tx}

\maketitle

\begin{figure*}[htbp]
\centerline{\includegraphics[width=1.9\columnwidth]{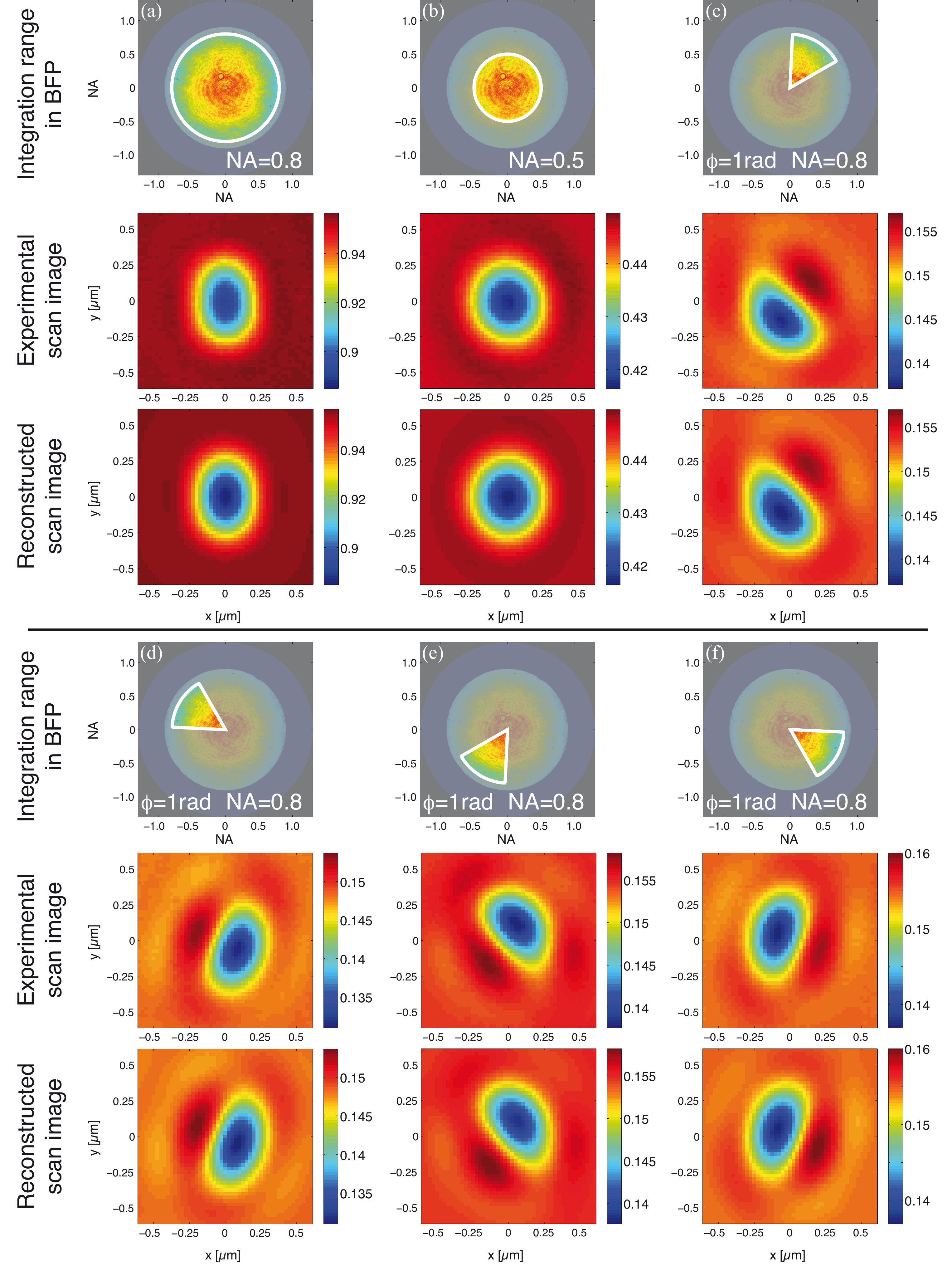}}
\caption{\label{fig1}(color). Experimentally achieved scan images of a tightly focused linearly (y-polarized) Gaussian beam for the angular ranges shown in the back focal plane of the collection microscope objective (first row), and comparison with the scan images derived from the reconstructed focal field distribution (last row).}
\end{figure*}

The experimental reconstruction of the focal field distribution of a tightly focused $y$-polarized Gaussian beam is realized with the setup schematically shown in Figure 3(a) of the main manuscript and described in detail in Ref.~\cite{Bauer2013}. A high numerical aperture (NA) objective with an NA of $0.9$ and a focal length of $f=1.65$ mm focuses the beam (beam width $w_0=2 \text{ mm}$, wavelength $\lambda=530 \text{ nm}$) to its diffraction limited focal spot. A single spherical gold nanoprobe with a radius of $42$ nm, immobilized on a glass substrate, is scanned step-wise through the focal plane with a lateral precision of approximately $2$ nm. The light transmitted and scattered into the forward direction is collected by an oil-immersion microscope objective ($\text{NA} = 1.3$), and its back focal plane (BFP) is imaged onto a camera. Integrating over different sectors of the BFP image for each position of the nanoprobe relative to the focal field distribution then allows for recording scan images corresponding to different collection angles. The resulting scan images for two integration sectors with an area spanning the full azimuthal angle around the optical axis and four angular sectors each spanning an azimuthal angle of $1 \text{ rad}$ are shown in Fig.~\ref{fig1}, with the chosen sectors displayed superimposed on a symbolic BFP image in the first row for each integration range. 

The integration over the intensity distribution in the BFP up to an NA of $0.8$ results in a scan image whose transmittance distribution is slightly elongated along the $y$-direction. This elongation is an indication of the expected longitudinal electric field component along the initial polarization direction due to the sensitivity of the collection angle to both the transverse and longitudinal electric field components. Reducing the integration radius to an NA of $0.5$ reduces the collection efficiency of the longitudinal electric field component with respect to the transverse components, leading to a nearly rotational symmetric scan image, as can be seen in Fig.~\ref{fig1}(b). An integration of the transmitted intensity over an angular sector rather than a full $2 \pi $ solid angle in the BFP of the collection objective (see Figs.~\ref{fig1}(c-f)) reveals the interferometric information necessary to reconstruct the full three-dimensional amplitude and phase distribution of the focal field. Here, the sector integrated scan images show a two-lobe pattern each, with the power in one lobe dropping below the collected background power. The orientation of the two lobes depends on the orientation of the angular sector in the back focal plane. The chosen sectors represent essentially four different observation directions of the interaction of the focal field with the nanoprobe.

The experimental far-field intensity data presented in Figure \ref{fig1} is used as input vector for the reconstruction algorithm described in detail in Ref.~\cite{Bauer2013}. The basic concept of the algorithm is the decomposition of the focal field distribution into vector spherical wave functions (VSWFs) \cite{Tsang2000} and the subsequent analytical expression of the scattering off the nanoprobe via generalized Lorenz-Mie theory, including the interaction with the air-glass interface.

The focal field distribution $\mathbf{E}_\text{in}$ can -- in this VSWF basis -- be written as
\begin{equation}
\mathbf{E}_\text{in}(\mathbf{r}) = \sum \limits_{n=1}^\infty \sum \limits_{m=-n}^n A_{mn} \mathbf{M}_{mn}(\mathbf{r}) + B_{mn} \mathbf{N}_{mn}(\mathbf{r}) \; ,
\end{equation}
with $\mathbf{M}_{mn}$ and $\mathbf{N}_{mn}$ representing regular VSWFs that can be associated with magnetic and electric multipoles, respectively, and are expanded around the geometrical focus of the beam \cite{Tsang2000}. With this decomposition the full focal field information is contained in the complex expansion coefficients $A_{mn}$ and $B_{mn}$. The interaction of the focal field distribution with the well characterized nanoprobe (here: spherical gold nanoparticle with radius $r=42 \text{ nm}$ and relative permittivity $\epsilon_\text{Au}=-3.4+2.4\imath$ at $\lambda = 530 \text{ nm}$) is then following classical scattering theory, where the transmission and reflection at the air-glass interface is taken into account using an effective scattering matrix $\hat{\mathbf{T}}_\text{eff}$ (see Ref.~\cite{Bauer2013} for details of this matrix). The total field transmitted in the forward direction can thus be calculated by
\begin{equation}
\mathbf{E}_\text{t} (\mathbf{r}) = \mathbf{E}_\text{in,t} + \mathbf{E}_\text{sca,t} = \left( 1 + \hat{\mathbf{T}}_\text{eff} \right) \mathbf{E}_\text{in,t} \; ,
\end{equation}
where $\mathbf{E}_\text{in,t}$ is the examined focal field and $\mathbf{E}_\text{sca,t}$ the field scattered off the nanoprobe, both transmitted trough the air-glass interface via Fresnel equations.
 
The angularly resolved optical power of the transmitted light ${P}_\text{t} (\theta, \phi)$ in the BFP of the collection objective can be written as:
\begin{align}
{P}_\text{t} (\theta, \phi ) &= P_\text{in} + P_\text{sca} + P_\text{ext} \\
P_\text{ext} &= \frac{1}{2}\text{Re} \left( \mathbf{E^*_\text{in}} \times \mathbf{H_\text{sca}} + \mathbf{E^*_\text{sca}} \times \mathbf{H_\text{in}} \right) \nonumber \; ,
\end{align}
with $P_\text{in}$, $P_\text{sca}$ and $P_\text{ext}$ representing the part of the optical power originating from the initial light field, the scattered field, and the interference between both fields, called extinction in classical Lorenz-Mie-theory. The magnetic field components of the scattered and incoming light are represented by $\textbf{H}$.

This angularly resolved detection of the optical far-field power allows for integrating over sectors with different effective observation directions, preserving the interference information about the relative phase between the different expansion coefficients and thus between the focal field components. By using a fixed cut-off of the multipolar expansion order (here: maximum order of $n=8$), it is possible to unambiguously invert the set of equations relating the far-field power collected in different angular sectors for all positions of the nanoprobe relative to the focal field to the complex VSWF expansion coefficients. Ultimately, this permits the reconstruction of the amplitude and phase distribution of all three focal electric field components.

\begin{figure*}[htbp]
\centerline{\includegraphics[width=2\columnwidth]{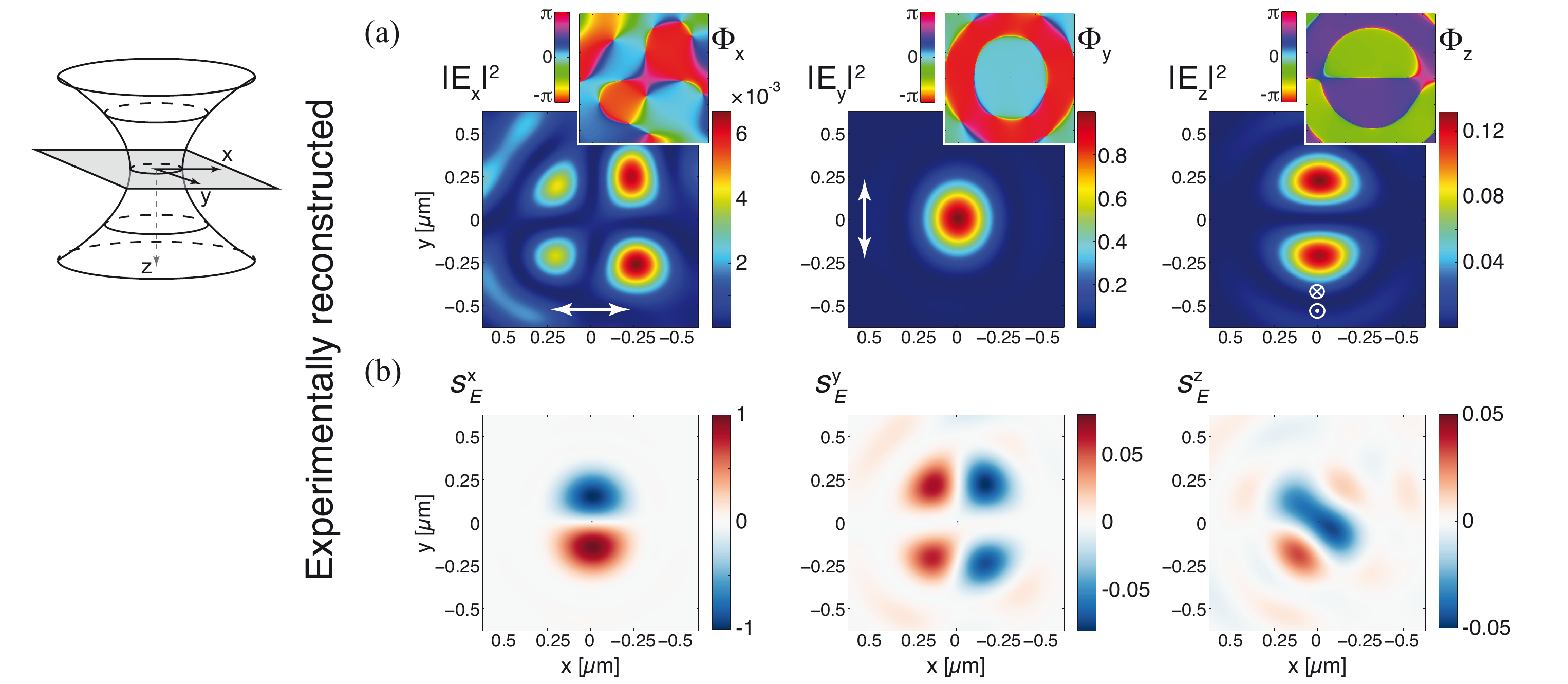}}
\caption{\label{fig2}(color). (a) Experimentally reconstructed electric energy density distributions $\lvert E_x \rvert^2$, $\lvert E_y \rvert^2$ and $\lvert E_z \rvert^2$, and their respective relative phase distributions $\Phi_x$, $\Phi_y$ and $\Phi_z$ (insets) in the focal plane. The energy density distributions are normalized to the maximum total electric energy density $\lvert \mathbf{E} \rvert^2$. (b) Components of the electric spin density ${s}_E^x$, ${s}_E^y$ and ${s}_E^z$, derived from the reconstructed electric field. The spin density distributions are normalized to the maximum total spin density $\lvert \mathbf{s}_E \rvert$.}
\end{figure*}

With the experimental scan images shown in Figure \ref{fig1} as input, the reconstruction algorithm leads to a set of multipolar expansion coefficients that can be used to calculate a set of reconstructed scan images using the same integration range as in the experiment  (see last row for each angular range in Fig.~\ref{fig1}), exhibiting an excellent overlap with the experimental input data. The experimentally determined focal field distribution of the electric field components shown in Figure \ref{fig2} can then be calculated from these reconstructed multipolar expansion coefficients.

These results also allow for determining the spin angular momentum density at each point in the focal plane using the relation
\begin{align}
\mathbf{s}_E=\frac{\epsilon_0}{4 \omega} \Im \left( \mathbf{E^*} \times \mathbf{E} \right) \; ,
\end{align}
already displayed as Eq.~(2) in the main manuscript. The resulting experimentally retrieved focal distributions are in good agreement with the distributions calculated by vectorial diffraction theory (\cite{L.NovotnyandB.Hecht2006}; see also Fig.~1 in the main manuscript). The non-zero longitudinal spin distribution present in the experimental field is caused by the non-planar phase profiles of the focal electric field components, which can be seen in the slight deformations of the $x$-components of the reconstructed electric field.

\begin{figure*}[htbp]
	\centerline{\includegraphics[width=2\columnwidth]{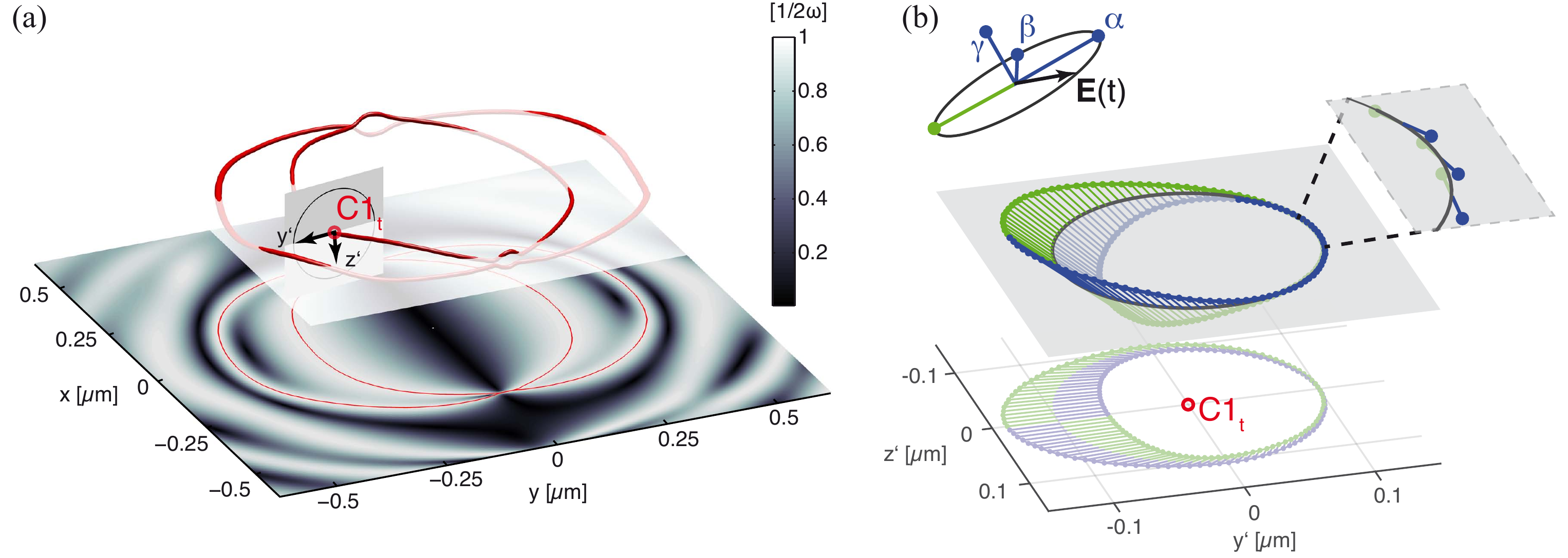}}
	\caption{\label{fig3}(color). (a) The experimentally reconstructed focal distribution of $\left|\textbf{s}_{E}\right| / w_E $ and the projections of the two innermost C-lines. The lower part depicts the projection of these C-lines onto the focal plane and the corresponding distribution of $\left|\textbf{s}_{E}\right| / w_E $.  The red lines in the upper part of (a) depict the 3D trajectories of both C-lines, which are crossing the focal plane (transparent white plane) repeatedly. The marked C-point $\text{C1}_\text{t}$ corresponds to the one considered in (b). (b) Optical polarization M\"obius strip with one half-twist, generated by tracing the major axis of the polarization ellipse around the C-point with purely tSAM at the intersection of the plane of observation marked in gray. The trace radius was chosen to be $100 \text{ nm}$. The magnified part of the trace shows that the major axis is, due to slight phase aberrations, at no time pointing along the trace direction. The occurring weak $x'$-component of the electric field is magnified 4 times to show the half-twist of the major axis of the polarization ellipse more distinctly.}
\end{figure*}

With the obtained decomposition of the reconstructed focal field distribution into VSWFs, the electric field can be retrieved not only in the focal plane but in a volume around the focus (see Fig.~3 in the main manuscript and Fig.~\ref{fig3}(a)). To investigate the optical polarization topologies present in this field around C-points associated with points of purely transverse spin density, the field in the focal volume can be examined for points with $\frac{\lvert \mathbf{s}_E \rvert}{w_E} = \frac{1}{2 \omega}$ (see Eq.~(3) in the main manuscript) and $s_E^z=0$. For the focal distributions shown in Figure \ref{fig2}, this set of conditions is fulfilled at the point $\text{C1}_\text{t}$ with coordinates $x=-100 \text{ nm}$, $y=-269 \text{ nm}$, and $z=3 \text{ nm}$ (see Fig.~\ref{fig3}(a)). The normal vector of the principal plane of the polarization ellipse at this point is thus located in the transverse plane, confirming the point to exhibit purely tSAM.

Tracing the major axis of the polarization ellipse $\boldsymbol{\alpha}(\mathbf{r})$ on a circle with radius $100 \text{ nm}$ around said C-point $\text{C1}_\text{t}$ results in the optical polarization M\"obius strip shown in Fig.~\ref{fig3}(b). Due to the non-planar phase front in all three electric field components in the transverse plane of the tightly focused light beam through this chosen C-point, the M\"obius-like polarization topology present in the numerically considered focal field distribution evolves to a generic M\"obius strip with one half twist, as can be seen by the inset in Fig.~\ref{fig3}(b). The major axis of the polarization ellipse at the trace point with $z'=0$ is tilted out of the principal plane of the C-point, allowing for unambiguously designating a twist index of $+1/2$ to this M\"obius strip.

\bibliography{bib}